\documentclass[fleqn,usenatbib]{mnras}

\usepackage{newtxtext,newtxmath}

\usepackage{graphicx}	
\usepackage{amsmath}	
\usepackage{amssymb}	
\usepackage{adjustbox}
\usepackage{threeparttable}
\usepackage{lipsum,cuted}
\usepackage{tabto}
\usepackage{float}
\usepackage{color}
\usepackage{amssymb} 
\usepackage{amsfonts,amsmath,amscd} 
\usepackage[subnum]{cases}
\usepackage{mathtools}
\usepackage{verbatim}
\usepackage{t1enc}
\usepackage[T1]{fontenc}
\usepackage{ae,aecompl}
\usepackage{graphics}
\usepackage{adjustbox}
\usepackage{caption}
\usepackage{eufrak} 
\usepackage{euscript} 
\usepackage[normalem]{ulem} 
\usepackage{makeidx} 
\usepackage[usenames,dvipsnames]{pstricks}
\usepackage{epsfig}
\usepackage{pst-grad} 
\usepackage{pst-plot} 
\usepackage{pstcol,times,pstricks,pst-node,pst-coil} 
\usepackage{multicol} 
\usepackage{multirow} 
\usepackage{indentfirst} 
\usepackage{natbib} 
\usepackage{setspace} 
\usepackage{threeparttable}
\usepackage{longtable}
\usepackage{pdflscape}

\newcommand{\Msun}{M_{\odot}}

\newcommand{\lppr}{\stackrel{<}{\scriptstyle \sim}}
\newcommand{\lappr}{\raisebox{-0.4ex}{$\lppr$}}
\newcommand{\bse}{{\sc bse }}

\title[Space density of CVs and period bouncers]
{No cataclysmic variables missing: higher merger rate brings into agreement observed and predicted space densities}
\author[Belloni et al.]
{Diogo Belloni$^{1,2}$\thanks{E-mail: belloni@camk.edu.pl (DB)},
Matthias R. Schreiber$^{3,4}$\thanks{E-mail: matthias.schreiber@uv.cl (MRS)}, M{\'o}nica Zorotovic$^3$\thanks{E-mail: mzorotovic@dfa.uv.cl (MZ)}, 
\newauthor Krystian I{\l}kiewicz$^1$, Jarrod R. Hurley$^5$, Mirek Giersz$^1$ and Felipe Lagos$^3$\\
$^{1}$ Nicolaus Copernicus Astronomical Centre, Polish Academy of Sciences, ul. Bartycka 18, PL-00-716 Warsaw, Poland \\
$^{2}$ CAPES Foundation, Ministry of Education of Brazil, DF 70040-020, Brasilia, Brazil \\
$^{3}$ Instituto de F{\'i}sica y Astronom{\'i}a, Universidad de Valpara{\'i}so, Av. Gran Breta{\~n}a 1111 Valpara{\'i}so, Chile \\
$^{4}$ Millenium Nucleus for Planet Formation, Universidad de Valpara{\'i}so, Valpara{\'i}so 2360102, Chile \\
$^{5}$ Centre for Astrophysics and Supercomputing, Swinburne University of Technology, Hawthorn, VIC 3122, Australia}

\begin{document}

\date{ Accepted 2018 May 29. Received 2018 May 4; in original form 2018 March 19}

\pagerange{\pageref{firstpage}--\pageref{lastpage}} \pubyear{2018}

\maketitle

\label{firstpage}
\begin{abstract}

The predicted and observed space density of cataclysmic variables (CVs) have been 
for a long time discrepant by at least an order of magnitude. The standard model 
of CV evolution predicts that the vast majority of CVs should be period bouncers, 
whose space density has been recently measured to be 
$\rho \lesssim 2 \times 10^{-5}$ pc$^{-3}$. 
We performed population synthesis of CVs using an updated version 
of the Binary Stellar Evolution ({\sc bse}) code for single and binary star evolution.
We find that the recently suggested empirical prescription of consequential angular 
momentum loss (CAML) brings into agreement predicted and observed space densities of 
CVs and period bouncers. To progress with our understanding of CV evolution it is 
crucial to understand the physical mechanism behind empirical CAML.
Our changes to the {\sc bse} code are also provided in details, which will allow the community 
to accurately model mass transfer in interacting binaries in which degenerate objects 
accrete from low-mass main-sequence donor stars. 
\end{abstract}


\section{INTRODUCTION}

Cataclysmic variables (CVs) are interacting
binaries composed of a white dwarf (WD) that accretes matter
stably from a low-mass main-sequence (MS) or slightly
evolved star \citep[e.g.][for a comprehensive review]{Knigge_2011_OK}.
CVs are the most numerous close compact binary stars and are therefore superb
systems to progress with our understanding of the long-term evolution of such
systems. Despite successfully explaining crucial characteristics of the
observed orbital period distribution of CVs, the standard model of CV
evolution that has been developed during the 
last decades fails when confronted with other observations of CVs.  

The most important feature observed 
in populations of CVs is the paucity of systems in the range of 2 hrs  
$\lesssim P_{\rm orb} \lesssim $  3 hrs,
known as the {\it orbital period gap} \citep[e.g.][]{Knigge_2006}. 
This orbital period gap is explained in the 
standard model for CV evolution by assuming disrupted magnetic braking
\citep[MB,][]{Rappaport_1982,spruit+ritter83-1,Howell_2001}. According to this
scenario, a CV above the gap is 
driven towards shorter
orbital periods by angular momentum loss (AML) owing 
to gravitational radiation (GR) and MB.
The latter is assumed to be orders of magnitudes 
stronger than the 
former at this stage and drives strong mass transfer. As a result of this strong mass 
transfer, the donor star is out of thermal equilibrium and significantly
bloated compared to isolated MS stars of the same mass. 
When the donor star becomes fully convective (at an orbital period of 
$\approx3$\,hrs), MB ceases, the mass transfer rate is reduced, 
and the donor star has time to relax to its thermal equilibrium radius. 
As a consequence, mass transfer completely stops and the CV converts 
into a close detached WD-MS binary until mass transfer starts 
again at an orbital period of $\approx2$\,hrs.
This explanation of the orbital period gap has been confirmed 
in several ways: \citet{Knigge_2006} showed that the radii of CVs at 
long orbital periods significantly exceed those of single MS stars; 
\citet{schreiberetal10-1} confirmed disrupted MB by 
showing that the evolutionary time-scale of post common envelope binaries (PCEBs)
significantly increases at the fully convective boundary; and finally
\citet{Zorotovic_2016} provided the most direct evidence by identifying a population 
of detached CVs crossing the gap. Explaining successfully the existence of the
orbital gap is a great achievement 
of the standard model for CV evolution.

The standard model also  explains reasonably well the existence of a {\it minimum
orbital period}, i.e. CVs do not appear to exist with orbital periods shorter than a
minimum value which is observationally determined to be at about 80 min \citep{Gansicke_2009}.  
In the context of the standard model, this 
can be explained as follows. 
After crossing the orbital period gap and restarting mass transfer again, 
the secondary is  
increasingly driven out of thermal equilibrium by mass loss. 
At some point, when the donor becomes strongly degenerate,  
its thermal time-scale exceeds
the mass loss time-scale and as a consequence it expands in response to the mass loss which
causes its orbital period to increase \citep[e.g.][]{King_1988}. The CVs with
substellar secondaries that  have passed the orbital period minimum 
are called period bouncers and according to the standard model they should
make up a large fraction of all CVs, e.g. \citet{Kolb_1993} predicted up to 70
per cent of all CVs should be period bouncers.

While the existence of a period gap and the period minimum are well explained
by the standard scenario of CV evolution, 
other predictions of the model dramatically
fail when compared to observations. 
The predicted space densities of CVs 
\citep[e.g.][]{Kool_1992,Kolb_1993} 
exceed those
derived from observations 
\citep[e.g.][]{Schreiber_2003,Pretorius_2012,Britt_2015}
by 1-2 orders of magnitude;
models predict a larger fraction of systems residing below the gap than
indicated by observations, and the WD masses expected in CVs are much smaller than those 
derived from observations \citep{Zorotovic_2011}.

It is well-known that AML in CVs determines their secular evolution. 
Apart from the alredy mentioned systemic AML (i.e. MB and GR), which
is independent of mass transfer, there is also AML that is a consequence of
mass transfer. This type of AML does not act if there is no mass
transfer and is therefore called {\em{consequential}} angular momentum loss \citep[CAML, e.g.][]{King_1995}.
Candidates for CAML are circumbinary disks \citep[e.g.][]{Willems_2005}, 
hydromagnetic accretion disk winds \citep[e.g.][]{Cannizzo_1988}, 
and AML associated with mass loss due to nova eruptions \citep{King_1995}.

Recently, \citet{Schreiber_2016} revised the standard model for CV
evolution and showed that if CAML in CVs is inversely proportional to the WD
mass, especially CVs with low-mass WDs run
into dynamically unstable mass transfer and the WD and its companion
merge. This can solve all three problems 
mentioned above (space density, WD mass, period
distribution).
In addition it seems that the merging CVs can explain the existence of single
low-mass WDs \citep{Zorotovic_2017}. 
Despite some previously suggested formation scenarios for these single low-mass WDs, e.g. 
(i) single star evolution with enhanced mass loss \citep[][]{Castellani_1993}; 
(ii) mass ejection by massive planets \citep[e.g.][]{Nelemans_1998};
(iii) supernova stripping \citep[e.g.][]{Justham_2009}; or
(iv) double helium-core WD merger \citep[e.g.][]{Saio_2000}, 
the idea of producing the single low-mass WDs from merging CVs is attractive as the predicted relative numbers seem to perfectly match the observations \citep{Zorotovic_2017}. 

The main mechanism thought to be responsible for the postulated dependence of
CAML on WD mass
are nova eruptions \citep{Schreiber_2016,Nelemans_2016}. 
Frictional AML produced by novae depends strongly 
on the expansion velocity of the ejecta \citep{Schenker_1998}, and 
for low-mass WDs, the expansion velocity is small \citep{Yaron_2005}. This may lead 
to strong AML by friction that makes most CVs with He WDs dynamically unstable
and the two stars merge instead of experiencing stable mass transfer. 
However, as the details of the physical mechanism behind the proposed
CAML prescription are not fully understood, 
this model is usually called empirical CAML (eCAML). 

Here we further compare the predictions of this revised empirical model for CV evolution
with observations. While \citet{Schreiber_2016} only compared the relative
numbers of CVs predicted by the classical and empirical CAML models, we here
determine the space density of CVs by calibrating the relative values using
the space density of single WDs.  
In particular, we also predict the space density of CVs with substellar donor stars
for both models.  
Recently, \citet{Santisteban_2017} estimated the space density of
period bouncers by searching in the Sloan Digital Sky Survey (SDSS) Stripe 82 data for 
drop-out eclipses in a sample of more than 2000 aparently single WDs
and found no such eclipses. This allowed them to infer an upper limit for
the period bouncer space density of $\lesssim 2 \times 10^{-5}$ pc$^{-3}$.

We find that the predicted space densities are in agreement with observations
if we assume eCAML while the classical model 
predicts more CVs than observed.

\section{{\sc bse} CODE}

The Binary Stellar Evolution ({\sc bse}) code developed by
\citet{Hurley_2000,Hurley_2002} is among the most frequently 
used codes to investigate secular evolution of CVs. 
It consists of 
a set of algorithms describing single and binary star
evolution. The main advantage of \bse is its speed and the  
generally high level of accuracy in the analytic fitting formulae 
on which it is based. 

Even though the {\sc bse} code is frequently used for population synthesis of 
CVs and related objects \citep[e.g.][]{Meng_2011,Zuo_2011,Schreiber_2016,Zorotovic_2017}
it is best suited to modeling just the early phases of their 
evolution, i.e. from the zero-age MS until 
the formation of PCEBs
\citep[e.g.][]{Schreiber_2016,Zorotovic_2016,Chen_2014}. This is mainly
because the {\sc bse} code in its original form 
includes only a simple description of the evolution of
accreting WD binary systems 
and comprehensive testing of degenerate mass-transfer phases was beyond the original scope. 
As such, fundamental ingredients 
of CV evolution are currently missing \citep[][see Section 5.2]{Belloni_2017}, 
and using \bse for CVs can thus 
lead to inaccurate predictions for e.g.  
mass transfer rates, duty cycles, or the orbital period and  
donor star mass distributions of CVs.

To overcome these shortcomings of the original {\sc bse} code, 
we updated the code in order to include state-of-the-art 
prescriptions for CV evolution. Our new version allows accurate modeling of 
interacting binaries in which degenerate objects are accreting 
from low-mass MS donor stars. In particular, this revised 
\bse code is suitable for performing state-of-the-art simulations of CV 
evolution while it remains fast enough to allow for large population synthesis.

In what follows we briefly discuss the major changes, and provide more details in 
the Appendix. The main upgrade of the code is a revision of the mass transfer 
rate equation which is now based on the model of \citet{Ritter_1988} 
and has been properly calibrated for CVs (Section \ref{mtr}). 
We have also added the radius increase/decrease of low-mass 
MS donors that is expected when mass transfer is
turned on/off (Section \ref{donor}), which is fundamental to reproducing the observed 
orbital period gap in CV populations.
We have also incorporated new options for systemic and CAML
 (Section \ref{aml}), which now includes the
MB prescription of \citet{Rappaport_1983} and CAML prescriptions described in
\citet{Schreiber_2016}.
Finally, we have adopted different stability criteria for dynamical and thermal mass 
transfer from MS donors (Section \ref{instcrit}), which now depend on the adiabatic 
mass-radius exponent and the mass-radius exponent of the MS star and on the assumed
CAML \citep{Schreiber_2016}.

In Section \ref{compoldnew}, we compare the predictions of the original \bse code with those obtained with
our upgraded version for the evolution of individual CVs, and show that 
our modifications lead to mass transfer 
rates as well as orbital periods and donor masses in much better agreement
with the observations.

\section{BINARY POPULATION MODEL}

We performed binary population synthesis using an initial population of one
million binary stars generated according to the following initial  
distributions: 
(i) The primary is obtained from the initial mass function (IMF) proposed by \citet{Kroupa_1993} in the range $[0.08,100]$ M$_\odot$;  
(ii) The secondary is obtained assuming a uniform mass ratio distribution, where $M_2 \leq M_1$, and requesting that $M_2\geq0.07$; 
(iii) The semi-major axis follows a log-uniform distribution in the range $[10^{-0.5},10^{4.5}]$ R$_\odot$. 
(iv) The eccentricity follows a thermal distribution in the range $[0,1]$. 

For the common-envelope phase (CEP), we assumed that only 25 per cent of the variation in the 
binary orbital energy  contributes to the expulsion of the
CE, and that the binding energy parameter is variable and properly calculated
assuming no contributions of recombination energy for expelling the CE.
This set of parameters is consistent with recent investigations that have concluded that 
WD-MS binaries experience strong orbital shrinkage during the CEP  
\citep[e.g.][]{Zorotovic_2010,Toonen_2013,Camacho_2014,Cojocaru_2017}.
We furthermore assume that the number of CVs that evolve through a  phase of thermal
time-scale mass transfer before becoming CVs is negligible. 
This is most likely a reasonable assumption as observations indicate that only 
$10-15$ per cent of all CVs form trough this channel \citep{Gaensicke2003}. 
However, it is important to stress that even our new version of the \bse code is unable to properly model thermal time-scale mass transfer
in CVs. This would require another modification of the prescriptions for mass transfer which is beyond the scope of the present paper. 
A recent attempt can be found in \citet{Wijnen_2015} who intended to model thermal time scale mass transfer phase with the binary\_c code \citep[e.g.][]{Izzard_2006}, which is similar to \bse, and concluded that a revision of the mass transfer equation was required. The fact that the revised \bse code presented here is unable to properly model thermal time scale mass transfer is clearly a flaw of the new code which should be carefully considered when analysing the distributions produced by the code.

As shown in \citet{Schreiber_2016}, the simulated population of CVs is 
strongly affected by the critical mass ratio that is defining the limit
between stable and unstable mass transfer. 
Apart from the intrinsic (often also called systemic)  AML due 
to GR and MB, CAML, i.e. AML due to mass transfer and mass loss during the nova eruptions,
is expected to play a key role in CV evolution \citep[][]{Schreiber_2016,Zorotovic_2017}.
We test here two different models for CAML: the classic model for CAML  
\citep[cCAML,][]{King_1995}, and the empirical CAML (eCAML), formulated by
\citet{Schreiber_2016} that recently has shown to solve several problems 
between model predictions and observations of CVs.
See Section \ref{caml} for additional details about both 
prescriptions.

As in \citet{Goliasch_2015}, we assumed that the IMF is constant in time 
and that the binary fraction is 50 per cent 
\citep[consistent with the binary fraction of WD primary progenitors, see][]{Patience_2002}.
In addition, we assumed a constant star formation rate 
\citep{Weidner_2004} during the lifetime
of the Galactic disk, which is assumed here to be $\approx$ 10 Gyr old. 
This way, the birth time of each binary and single star in the simulations
is chosen randomly, assuming a uniform distribution, and they are evolved 
from the birth time until 10 Gyr.  

In order to compute the CV space density, we follow the procedure described
in \citet[][see their sections 2.2.3 and 2.2.4]{Goliasch_2015}. As we simulate both
single stars and binaries,  
we can normalize the results of our population synthesis such that the number
of single WDs corresponds to a specific birthrate of WDs in the Galactic
disk. We adopt a WD formation rate of 0.4 WD yr$^{-1}$, which gives a total
number of $4 \times 10^9$ WDs in the disk. 
The total number of CVs was scaled the same way to obtain absolute numbers 
of systems that should be present
in the Galactic disk. Finally, we compute the space density by assuming
a Galactic volume of $5\times10^{11}$ pc$^3$ \citep[e.g.][]{Toonen_2017}.

\begin{table*}
\centering
\caption{Number and space density of present-day single WDs, CVs and period bouncers 
in our simulations, according to both models of CAML assumed here. The number of single WDs
corresponds to the total amount of single WDs formed in both single and binary evolution.
The number of CVs refers to the total amount of all types of CVs, i.e. long-period, gap CVs, short-period and
period bouncers. The numbers presented in columns labeled {\em{Modelling}} corresponds to the number obtained in
our simulations, while those in the columns labeled {\em{Absolute}} are scaled
with respect to the single WD birth rate.
Finally, the space density ($\rho$) was computed based on the absolute numbers, assuming a Galactic volume
of $5\times10^{11}$ pc$^3$.}
\label{Tab01}
\begin{adjustbox}{max width=\linewidth}
\noindent
\begin{threeparttable}
\noindent
\begin{tabular}{l|c|c|c|c|c|c|c|c|c|c|c|c}
\hline
\hline
CAML & \multicolumn{3}{c}{Single WDs} & & \multicolumn{3}{c}{CVs} & & \multicolumn{3}{c}{Period bouncers} \\
\hline
 & Modelling  & Absolute & $\rho$ [pc$^{-3}$] &  & Modelling  & Absolute & $\rho$ [pc$^{-3}$] & & Modelling  & Absolute & $\rho$ [pc$^{-3}$]  \\ 
\hline
classical & $95525$ & $4\times 10^9$ & $8\times 10^{-3}$ & & $1068$ & $4.472\times 10^7$ & $8.9\times 10^{-5}$ & & $717$ &  $3.002\times 10^7$ & $6.0\times 10^{-5}$ \\ \hline
empirical & $95979$ & $4\times 10^9$ & $8\times 10^{-3}$ & & $ 237$ & $0.988\times 10^7$ & $2.0\times 10^{-5}$ & & $181$ &  $0.754\times 10^7$ & $1.5\times 10^{-5}$ \\ \hline
\hline 
\end{tabular}
\end{threeparttable}
\end{adjustbox}
\end{table*}


\section{Results}
\label{results}

We ran population models of CVs using two prescriptions for
CAML and determined the predicted space densities
for all CVs and for period bouncers. 
Table \ref{Tab01} exhibits the number and space 
density of present-day single WDs, CVs 
and period bouncers predicted by our simulations.

\begin{figure*}
   \begin{center}
    \includegraphics[width=0.45\linewidth]{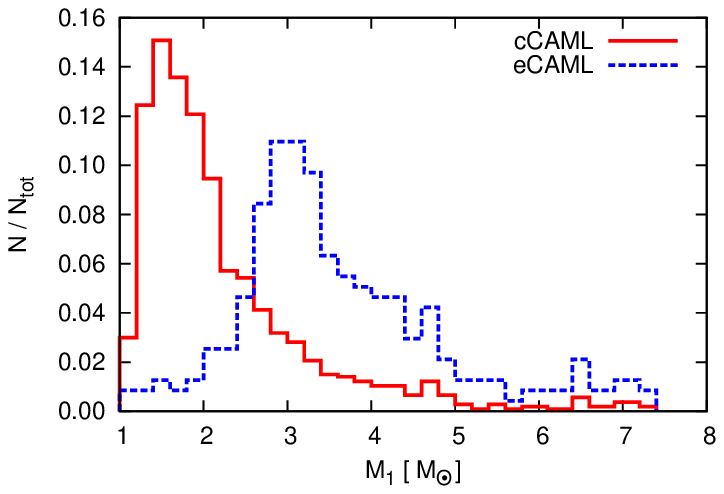} 
    \hspace{0.5cm}
    \includegraphics[width=0.45\linewidth]{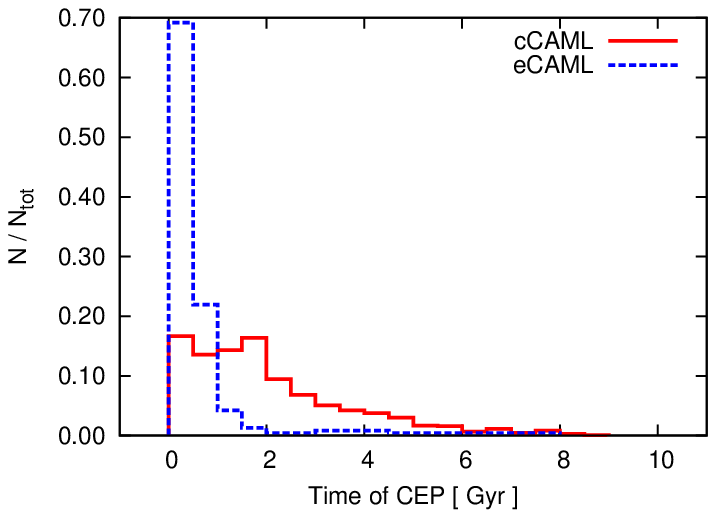} 
    \includegraphics[width=0.45\linewidth]{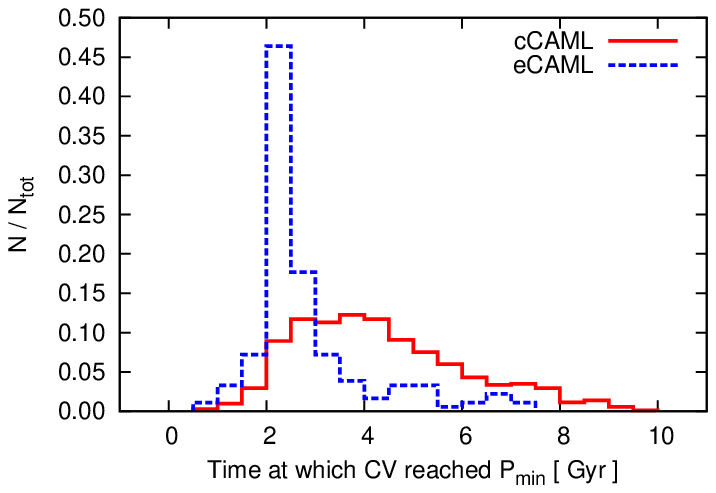} 
    \hspace{0.5cm}
    \includegraphics[width=0.45\linewidth]{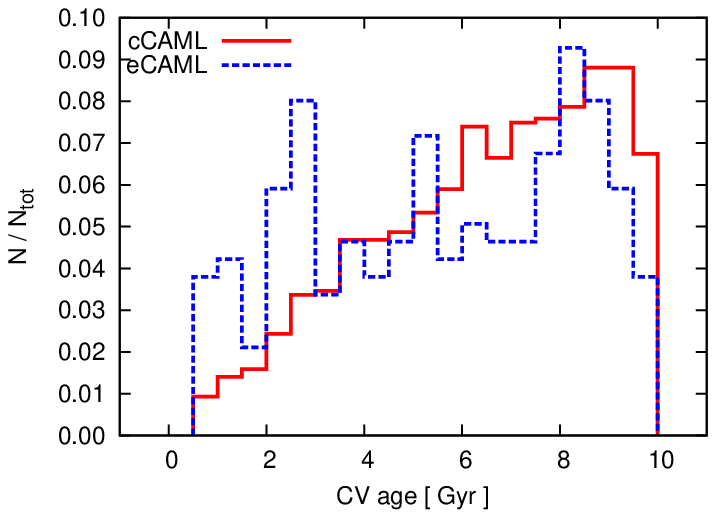} 
    \end{center}
  \caption{Distributions of all CVs in our simulations:
WD progenitor mass (top left-hand panel),  
time at which the CEP takes place (top right-hand panel), 
time at which CVs reach the period minimum  (bottom left-hand panel),
and CV age (bottom right-hand panel).}
  \label{Fig02} 
\end{figure*}

\begin{figure*}
   \begin{center}
    \includegraphics[width=0.45\linewidth]{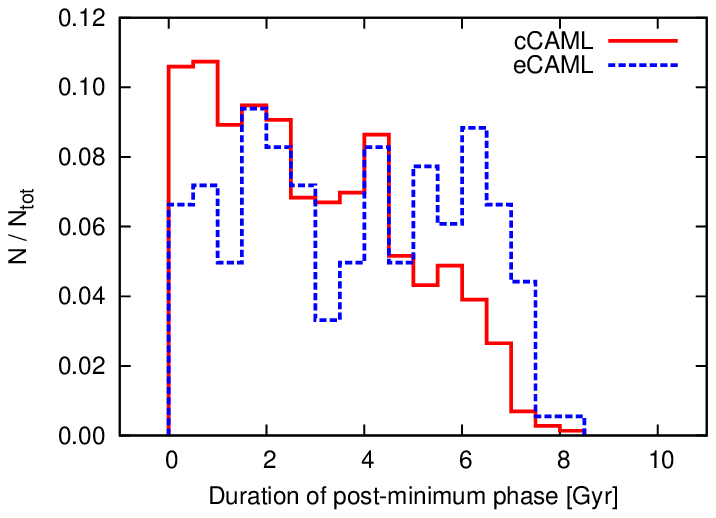} 
    \hspace{0.5cm}
    \includegraphics[width=0.45\linewidth]{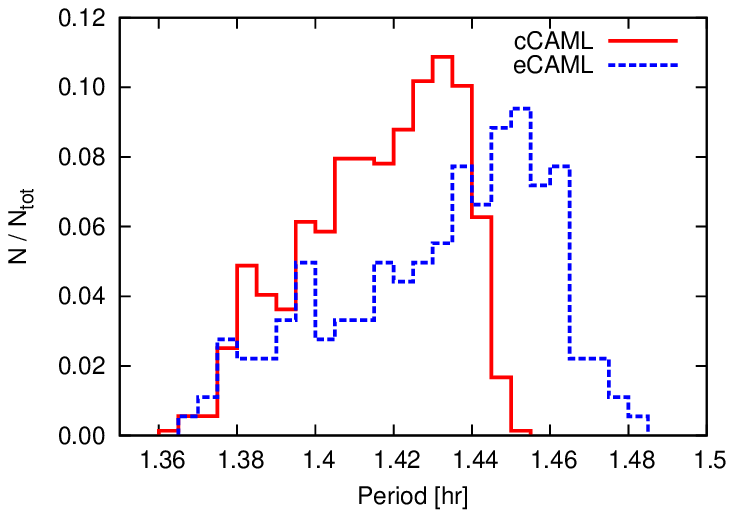} 
    \includegraphics[width=0.45\linewidth]{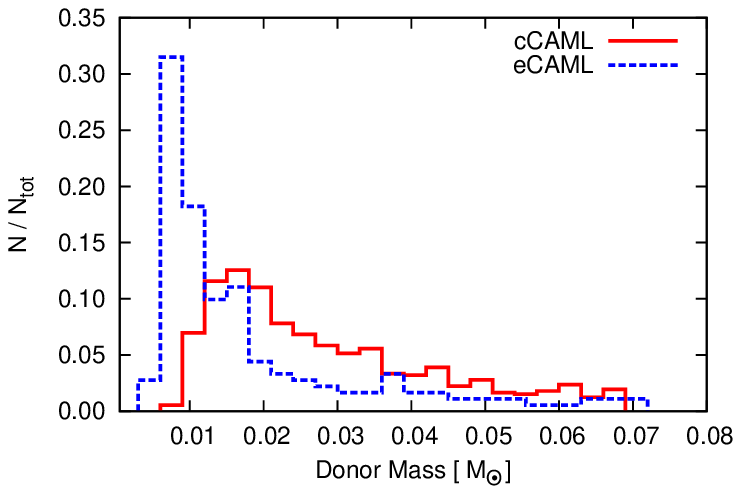} 
    \hspace{0.5cm}
    \includegraphics[width=0.45\linewidth]{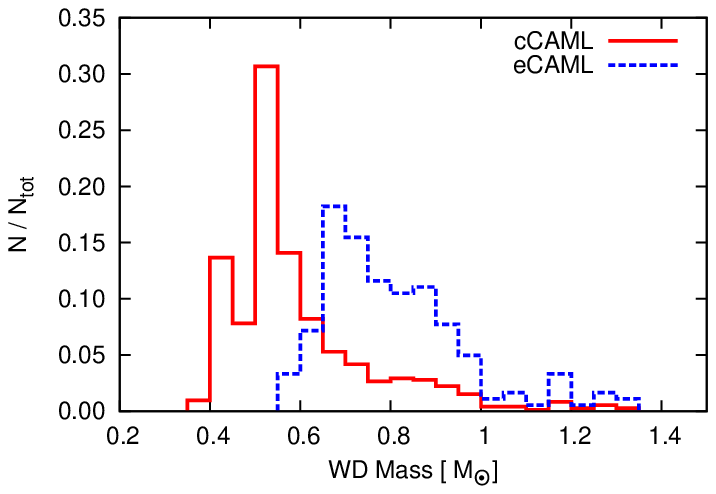} 
    \end{center}
  \caption{Distributions associated with period bouncers in our simulations:
lifetime as period bouncers (top left-hand panel),
orbital period (top right-hand panel),  
donor mass (bottom left-hand panel),
and WD mass (bottom right-hand panel).}
  \label{Fig03} 
\end{figure*}

The numbers of single WDs 
formed in our simulations, for
both cCAML and eCAML, are similar, although the number in eCAML is slightly 
larger than in cCAML due to the fact that the critical mass ratio for stable mass transfer
defined by eCAML leads to a narrower stable region in the parameter space of CVs
\citep[][see, e.g. figs. 1 and 2]{Schreiber_2016}, while most CVs whose
primaries are low-mass WDs merge, leading to more 
single WDs \citep{Zorotovic_2017}.
The same effect  significantly reduces the CV space densities predicted by
eCAML ($\approx2\times10^{-5}$ pc$^{-3}$) compared to the classical model ($\approx9\times10^{-5}$ pc$^{-3}$). 
As a much larger number of systems suffer dynamical unstable mass transfer,
less systems experience stable AML-driven mass transfer. 
 
Inspecting Table \ref{Tab01} in more detail, we find that the reduction in space
density due to the implementation of eCAML is smaller for period bouncers than 
for the entire CV population. Therefore, the space density of the total CV
population is somewhat less reduced than in \citet{Schreiber_2016} who 
ignored period bouncers with secondary masses below $0.05\Msun$. 
This different effect of eCAML on the space densities of the entire CV
population and period bouncers is caused by the mass dependence of
the additional AML in the eCAML scenario
as illustrated in Fig. \ref{Fig02}.

By postulating that systems with low-mass WDs preferentially merge
due to unstable mass transfer
(for a visualisation of this effect in the WD mass vs. mass ratio plane see fig. 1 of \citealt{Zorotovic_2017}), 
the WD mass distribution of CVs predicted by eCAML contains
only relatively massive WDs and these must have had quite massive MS
star progenitors (upper left panel). 
The intermediate-mass MS stars have short MS lifetimes and the
CEP is on average reached earlier than in the case of the classical model
(upper right).   
The shorter MS lifetimes for the CV primary 
progenitors in the eCAML model then translate into shorter 
times until a CV reaches the orbital minimum (lower right panel). The shorter
CV formation times in eCAML finally imply that this 
model predicts a flatter age
distribution of CVs than the classical 
model (bottom right panel). Consequently, the relative fraction of CVs that
are period bouncers is even larger in the eCAML model ($76$ per cent) than according to the
standard theory ($67$ per cent)

The different evolutionary time-scales also affect the 
properties of the population of period bouncers as shown in Fig.\,\ref{Fig03}. 
According to eCAML significantly more CVs have passed the period 
minimum more than $\sim\,5$\,Gyrs ago (top left-hand panel) and have therefore
reached longer orbital periods (top right-hand panel) and lower secondary star masses (bottom left-hand panel). 
The WD mass distributions of the period bouncers (bottom right-hand panel)
are similar to those of the entire CV population: 
while eCAML predicts only massive WD
systems to become CVs, the distribution predicted 
by the classical model is dominated by low-mass WDs, which is 
inconsistent with observations.  

One might think that the different evolutionary time scales and the different 
WD masses predicted by the two models translate into a
significant difference in the temperatures of the accreting WDs. 
In order to test if this is the case, 
we determined the effective temperature distributions of the predicted period
bouncer populations by calculating both the cooling temperature of the WD and
the temperature according to compressional heating. The former were determined
as in \citet{Zorotovic_2017},
i.e. via interpolation of DA (pure hydrogen atmosphere) WD evolutionary 
models by \citet{Althaus_1997}, for helium-core WDs, and by \citet{Fontaine_2001}, 
for carbon/oxygen-core WDs.
The temperature generated by compressional heating was determined from Eq.\,2 in
\citet{Townsley_2009}, i.e.

\begin{equation}
\frac{T_{\rm eff}}{{\rm K}} \ = \ 1.7\times10^4 \; \left( \, \frac{\langle \, \dot{M}_d \, \rangle}{10^{-10} \, {\rm M}_\odot \, {\rm yr}^{-1}} \, \right)^{0.25} \; \left( \, \frac{M_{\rm WD}}{0.9 \, {\rm M}_\odot} \, \right), \label{compheat}
\end{equation}
where $T_{\rm eff}$, $\langle \dot{M}_d \rangle$ and $M_{\rm WD}$ are the
WD effective temperature due to compressional heating, the average mass transfer
rate and the WD mass, respectively.

\begin{figure}
   \begin{center}
    \includegraphics[width=0.95\linewidth]{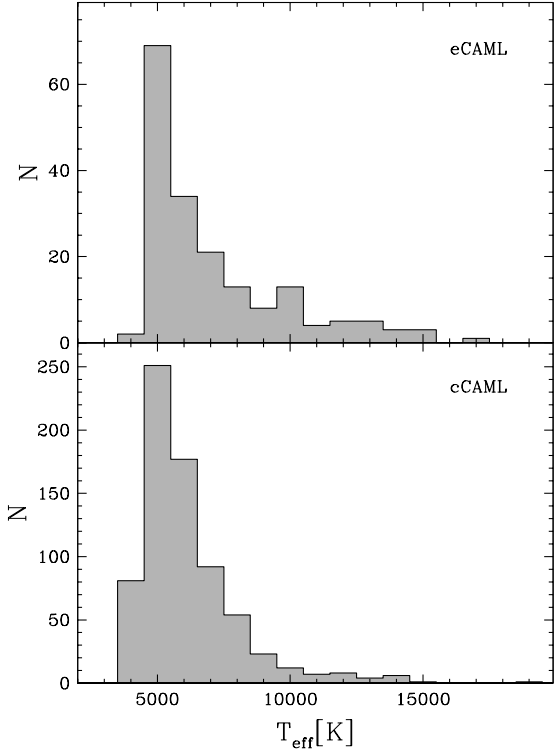}
    \end{center}
  \caption{The predicted distribution of WD temperatures in period bouncers
    computed by comparing the cooling temperature of the WD and the temperature
    derived from compressional heating and selecting the higher value for each WD. 
Most WD temperatures, in both models, lie between 4000 and 7000\,K.}
  \label{Fig04} 
\end{figure}

We find that in most cases, the compressional heating temperature exceeds   
the cooling temperature and therefore, based on Eq. \ref{compheat}, 
the mass transfer rate and the WD mass determine the WD effective temperature 
of most WDs in period bouncers. The
resulting distributions are shown in Fig.\,\ref{Fig04}. Apparently, both
distributions are very similar. We find that this is because the effects of the higher WD masses and the larger ages (and therefore smaller mass transfer rates) predicted by eCAML
cancel out each other: 
according to Eq.\,\ref{compheat} higher WD masses lead to higher WD effective temperatures while the smaller accretion rates in systems that passed the period minimum long time ago cause the compressional heating temperature to decrease.
For both models we therefore find that the predicted population of period bouncers is dominated by systems containing very cool WDs ($\sim\,5000\,$K) which
implies that the majority of these systems are very hard to find.

\section{Discussion}
We performed binary population synthesis of CVs  
and calculated the space densities of CVs and period bouncers
according to the standard model of CV evolution and the new empirical CAML
model suggested by \citet{Schreiber_2016}.

Our simulations involve several model assumptions, such
as a small CEP efficiency, flat initial mass ratio distribution,
log-uniform semi-major axis distribution and constant star formation rate.
Evidence is growing towards a low CEP efficency
\citep[e.g.][]{Zorotovic_2010,Toonen_2013,Camacho_2014,Cojocaru_2017} and constant
star formation rate 
\citep[e.g.][]{Kroupa_2013,Recchi_2015,Schulz_2015} but 
the situation with respect to initial binary distributions is less clear.

There seem to exist correlations amongst orbital parameters for different
stellar spectral types \citep[e.g.][]{DM_1991,Moe_2017} and assuming uncorrelated
distributions as done here might not be realistic.
In the particular case of WD progenitors, \citet{Rosa_2014}, 
by combining adaptive optics imaging and a multi-epoch common proper motion search,
found for A-type stars that the semi-major axis distribution forms a log-normal distribution
and the mass ratio distribution of closer ($a<$125 AU) binaries is distinct from that of 
wider systems, with a flat distribution for close systems and a distribution that tends 
towards smaller mass ratios for wider binaries.
Furthermore, \citet{Moe_2017} investigated binaries whose primaries are mainly A, B and O-type stars,
and after combining the samples from various surveys and correcting for their respective selection effects,
these authors concluded that the distributions are not independent and fitted joint probability
density functions $f(M_1,q,P,e) \neq f(M_1)f(q)f(P)f(e)$ to the corrected distributions. 
These fitted joint distributions are probably the most realistic available to the community and
should be incorporated into future binary population synthesis studies. 
In this paper we present the
impact of different CAML prescriptions on the predicted space densities and 
therefore preferred to assume uncorrelated distributions for comparison with previous
works. However, we plan to
investigate the impact of more complex initial distributions on the results of
binary population models in a future paper.

The values we obtain for the space density of CVs by assuming cCAML and eCAML
are $\approx~0.9~\times~10^{-4}$~pc$^{-3}$ and $\approx~2~\times~10^{-5}$~pc$^{-3}$,
respectively, i.e. the CV space density predicted by the eCAML 
model is considerably smaller than the value predicted by the 
classical model because a large number of systems merge instead of becoming CVs. 

Previous theoretical estimates of the CV space density 
cluster around mid-plane 
space densities of $10^{-4}$~pc$^{-3}$ and are thus in general agreement with
the value we found assuming cCAML. 
Using the CV formation rate 
derived by \citet{Kool_1992}, assuming an age of the Galaxy of $10$\,Gyrs 
and assuming that the lifetimes of CVs exceed this value, the predicted space density
is $(0.5-2)\times10^{-4}$~pc$^{-3}$.  
Normalizing the results of the binary population model for CVs from \citet{Kolb_1993} 
with the formation rate of single WDs, 
the predicted mid-plane CV
space density is $1.8\times10^{-4}$~pc$^{-3}$. In both studies the classical CAML
prescription has been assumed.
A significantly smaller value has been recently 
predicted by \citet{Goliasch_2015} who found 
a space density of $1.0(\pm0.5)\times10^{-5}$pc$^{-3}$. 
However, these authors did not describe which criterion has 
been used to separate
stable and unstable mass transfer and which form of 
CAML has been assumed in their calculations. 
It is therefore impossible to understand the significant difference
to other previous predictions.

The smaller space density we obtain 
assuming eCAML is in very good 
agreement with the space densities derived from 
observations.
For example, using the ROSAT 
North Ecliptic Pole survey  
\citet{Pretorius_2007} obtained  
a space density of the CV population 
of $1.1^{+2.3}_{-0.7}\times10^{-5}$pc$^{-3}$ and \citet{Hertz_1990}
derived $\approx(2-3)\times10^{-5}$pc$^{-3}$ using the 
Einstein Galactic Plane Survey.  
\citet{Schreiber_2003} derived a lower 
limit of $\sim10^{-5}$pc$^{-3}$ using a 
relatively small sample of PCEBs, the
progenitors of CVs, and estimating their evolutionary 
lifetimes. 
This lower limit, however, is based on the 
cCAML assumption and would become much smaller if 
eCAML applies as many of the considered PCEBs would merge when mass transfer
starts instead of becoming CVs.  
More recently, \citet{Britt_2015} 
derive a space density of long-period CVs 
of $5.6(\pm3.9)\times10^{-6}$pc$^{-3}$ using data from American Association of
Variable Star Observers. 
A more comprehensive review of 
observational estimates of 
the space density of CVs can be found in 
\citet[][their Fig.\,8]{Santisteban_2017}. 
As the values derived from observations cluster 
around $\sim10^{-5}$pc$^{-3}$, agreement with 
model predictions can be reached if 
eCAML is assumed while the classical models 
predict values significantly exceeding those derived 
from observations.  

An additional test for the different model predictions 
is offered by a recent measurement of an upper limit 
for the space density of CVs that passed the
period minimum \citep{Santisteban_2017}.
This derived upper limit is relatively uncertain
as it depends significantly on the assumed 
characteristic WD temperature
\citep[][see their fig.\,7]{Santisteban_2017}. 
However, according to our models, realistic characteristic WD 
temperatures should lie in the range of $4000-7000$\,K 
for both models (see Fig.\,\ref{Fig04}). Together 
with the different characteristic WD masses (0.6~M$_\odot$ for cCAML and 0.8~M$_\odot$ for
eCAML) and using Fig.\,7 of \citet{Santisteban_2017} this translates into  
upper limits derived from observations of $\sim\,1-10\,\times10^{-4}$ pc$^{-3}$ (eCAML) 
and $\sim\,4-60\,\times10^{-5}$ pc$^{-3}$ (cCAML) for the space density of
period bouncers. 

The upper limit derived from observations is similar to the model predictions
in the case of cCAML ($\approx6\times10^{-5}$~pc$^{-3}$) while it largely
exceeds the prediction in the case of eCAML
($\approx1.5\times10^{-5}$~pc$^{-3}$). 
This means that the observational constraints on the space density of period
bouncers currently available 
are not sufficient to distinguish between the two models. 
Deeper surveys and/or with shorter cadence are required to test models for CV
evolution. 

However, while we cannot exclude the classical model 
based on the upper limit for period bouncers alone, taking into account the measured and
predicted space densities of the entire CV population, 
it currently seems that the predictions of 
the eCAML model agree significantly better with the observations.   
We therefore predict that future surveys of period bouncers may find less
systems than expected from the classical model because according to eCAML 
there are less period bouncers and those that exist have
relatively long periods, low-mass companions, small accretion rates, and massive WDs. 
All these factors make their discovery even more challenging than
the population predicted by the classical model.


\section{Conclusion} 

We carried out population synthesis 
of cataclysmic variables (CVs) and the population of 
period bouncers among them and predicted space
densities for both populations.  
We considered two prescriptions 
of CAML, namely the classical non-conservative and 
the recently suggested empirical model. 
The latter is a good candidate 
to solve several problems related to CV evolution, like the missing low-mass
WDs in CVs or the period distribution 
\citep[see][for more details]{Schreiber_2016},
and may also provide an explanation for the 
existence of 
single He-core WDs \citep{Zorotovic_2017}.

We confirm here that the space densities of CVs and period bouncers predicted
by eCAML agree significantly 
better with the space densities derived from
observations than the classical CV evolution model. 
However, eCAML is a purely 
empirical model and as an important next step towards 
a global understanding of CV evolution we need to 
investigate the physical origin of 
the enhanced CAML for systems with low-mass 
WDs. As suggested by \citet{Schreiber_2016} 
and \citet{Nelemans_2016} this origin might 
be found by studying the impact of nova eruptions and 
possible frictional AML on the 
secular evolution of CVs.

\section*{Acknowledgements}

DB was supported by the CAPES foundation, Brazilian Ministry of Education through the grant BEX 13514/13-0 and by the National Science Centre, Poland, through the grant UMO-2016/21/N/ST9/02938. 
MRS acknowledges financial support from FONDECYT grant number 1141269. 
MZ acknowledges support from CONICYT PAI (Concurso Nacional de Inserci\'on en la Academia 2017, Folio 79170121) and CONICYT/FONDECYT (Programa de Iniciaci\'on, Folio 11170559)
KI  has  been  financed  by  the  Polish  Ministry  of  Science  and  Higher  Education  Diamond  Grant Programme via  grant  0136/DIA/2014/43.
MG was partially supported by the National Science Centre, Poland, through the grant UMO-2016/23/B/ST9/02732.

\bibliographystyle{mnras}
\bibliography{references}

\bsp


\appendix

\section{UPGRADES  TO THE  {\sc bse} CODE FOR POPULATION SYNTHESIS OF CVS}
\label{upgrades}

We describe here the main upgrades to the {\sc bse} code which
include a new mass transfer rate equation when the donor
is a Roche lobe underfilling star (Section \ref{mtr}), 
a new method for the response of low-mass MS donor stars to mass transfer
(Section \ref{donor}), new AML 
prescriptions (Section \ref{aml}), and new instability criteria for thermal and 
dynamical time scale mass transfer (Section \ref{instcrit}).
The upgrades we applied to \bse bring CV evolution and observations into agreement.


\subsection{Mass transfer onto degenerate objects}
\label{mtr}

In the original version of the {\sc bse} code, \citet{Hurley_2002} modelled the
mass transfer rate using an equation that depends on the fraction
of the donor that is overfilling its Roche lobe  ($R_{d}/R_{{\rm L},d}$), i.e. 
\begin{equation}
\dot{M}_{d} \ = \ f(M_d) \ [ \ \ln(R_{d}/R_{{\rm L},d}) \ ]^3 \ {\rm M}_\odot \ \rm{yr}^{-1}, \label{EQ1}
\end{equation}
where $R_d$ is the donor radius, $R_{{\rm L},d}$ is the donor Roche lobe radius, $M_d$ is the donor mass,
and the factor $f \ = \ 3 \times 10^{-6} \ [\min(M_d,5.0)]^2$ comes from numerical
experiments \citep[see also][]{Whyte_1980}. 
Despite some recent improvements \citep[e.g.][]{Claeys_2014,Wijnen_2015},
this prescription leads to inaccurate mass transfer rates if mass transfer 
is driven by AML such as in CVs and low-mass X-ray binaries. 
This is because Eq.\,\ref{EQ1} does not account for 
the structure of the donor star. In particular, it neglects the finite 
scale height of the donor's atmosphere.
In order to provide a mass transfer equation that works for CVs 
and related objects, we follow 
\citet{Ritter_1988} who considered that stars have extended 
atmospheres such that Roche lobe overflow can occur 
even if $R_{d} < R_{{\rm L},d}$. In this case we use the following 
equation
\begin{equation}
\dot{M}_{d} \ = \ \dot{M}_{0} \ \times  \ \exp \left( - \ \frac{R_{{\rm L},d} - R_{d}}{H_P} \right) \  \ {\rm M}_\odot \ \rm{yr}^{-1}, \label{EQ2}
\end{equation}
where $\dot{M}_{0}$ is the mass transfer rate when the donor fills exactly its Roche lobe and $H_P$ is the pressure scale height at the 
donor's photosphere. The pressure scale height at the donor's 
photosphere is computed 
based on Eq. 8b in \citet{Ritter_1988}, i.e.
\begin{equation}
H_P \ = \  \frac{\Re \, T_{d} \, R_{{\rm L},d}^2}{\mu \, G \, M_{d}}, \label{EQ_HP}
\end{equation}
where $T_{d}$ is the donor's effective temperature, $\Re$ is the gas constant,
$\mu$ is the mean molecular weight, and $G$ is the gravitational constant. In
addition, we calculate $\dot{M}_{0}$ according to
\begin{equation}
\dot{M}_{0} \ = \ \frac{2 \, \pi}{\sqrt{2.71828}}  \ \left( \frac{\Re \, T_{d}}{\mu} \right)^{3/2} \ \frac{R_{{\rm L},d}^3}{G \, M_{d}}  \ \rho_{ph}  \ F(q), \label{EQ_M0}
\end{equation}
where  $\rho_{ph}$ is the density in the donor's photosphere
and $F(q)\simeq1.23+0.4\log q$ 
\citep[see appendix of][]{Ritter_1988},
where $q = M_d/M_a$ is the mass ratio and $M_a$ is the accretor mass. 
In order to calculate $\rho_{ph}$ we employed MARCS model atmospheres
\citep{marcs_2008} and created a grid of 
densities at an 
optical depth of 2/3 ($\rho_{ph}$) 
for different $T_{d}$, donor's surface gravities, 
and metallicities. 
The resulting grid was interpolated to obtain $\rho_{ph}$ for a given donor 
star. In cases where the donor becomes strongly degenerate 
($M_d \lesssim 0.085$ M$_\odot$) MARCS model atmospheres 
become inaccurate 
and we thus use 
$\dot{M}_{0} = 6.4 \times 10^{-9}$ M$_\odot$ yr$^{-1}$ 
and $H_P = 10^{-4} R_d$ \citep[][table A1]{Ritter_1988}.
Equation~\ref{EQ2} has been largely used to calculate mass transfer rates 
of CVs \citep[e.g.][]{Davis_2008,Knigge_2011_OK,Zorotovic_2016} 
and accurately describes the physical processes occurring 
when mass transfer is turning on or off. In addition, 
mass transfer rates in CVs calculated according to 
Eq.~\ref{EQ2} generally agree with those derived from observations (depending on the assumed magnetic braking prescription, of course).

According to equation\,\ref{EQ2} Roche-lobe overflow starts when the extended atmosphere of the secondary reaches the Roche-radius, i.e. when the ratio between photospheric radius and Roche-radius is still smaller than 1. As explained in detail in \citet{Davis_2008}, calculating proper mass transfer rates from Eq.\,\ref{EQ2} requires to use relatively small time steps 
($\lappr\,10^{3-4}$ yrs) 
to keep the filling factor below 1.00001 as otherwise the mass transfer rates become unrealistically high.
This makes the revised {\sc bse} code somewhat slower, as we will discuss 
in Section \ref{finalc}.
Finally, we note that Eq. \ref{EQ2} is only applied in the upgraded 
code for binaries in which a degenerate object 
(WD, neutron star or black hole) is accreting
from either MS or giant stars underfilling their Roche lobe. Otherwise,
Eq. \ref{EQ1} is used to estimate the mass transfer 
rate.

\subsection{Donor expansion due to mass transfer from low-mass MS stars}
\label{donor}

With the new mass transfer algorithm, \bse reproduces  
the mass transfer 
rates predicted by other population models of CVs and those derived 
from observations of CVs. 
However, even with this new prescription for the mass
transfer rates, \bse is not able to reproduce
the most important feature in the observed orbital period distribution 
of CVs, i.e. the paucity of systems with orbital
periods in the range of $2-3\,$hrs. 

In order to simulate CV evolution with \bse we do not only need to 
upgrade the mass transfer algorithm as described in the previous section, 
we also need to incorporate the radius increase of low-mass MS 
donor stars as they respond to strong mass loss. 
The amount by which mass losing low-mass MS donors are bloated 
relative to an isolated star of the same mass depends on how much their 
thermal time-scale exceeds the mass loss time-scale. 
\citet{Knigge_2011_OK} derived radii of CV donors from observations of 
a relatively large sample of CVs and found CVs above the orbital period gap
to be bloated by about 30 per cent. 
In the original {\sc bse} code, the donor stars are not inflated which
results in incorrect evolutionary tracks for CVs. Most importantly, 
without the increased radii of donor stars above the gap, 
it is impossible to
reproduce the orbital period gap in binary population synthesis calculations. 
We include this fundamental ingredient of CV evolution in 
the upgraded {\sc bse} code by following \citet{Wijnen_2015}, i.e. we 
increase the donor star radii as soon as mass transfer starts 
until it reaches the
value derived from observations \citep{Knigge_2011_OK}.  

More specifically, the procedure is as follows. 
Let us consider first CVs that are born above the gap 
($M_d \geq 0.35$ M$_\odot$). For these we define a factor that 
regulates the inflation from the equilibrium radius to the CV donor star 
radius as 
\begin{equation}
f_{\rm c} \ = \ \frac{R_{d, {\rm CV}}^{\rm above}}{R_{d,{\rm eq}}}, \label{EQ10}
\end{equation}
where $R_{d, {\rm CV}}^{\rm above}~=~0.293(M_d/0.2)^{0.69}$ 
\citep{Knigge_2011_OK}
is the CV donor radius above the gap and $R_{d, {\rm eq}}$ is the equilibrium 
radius computed by {\sc bse} in the standard way.
In order to have a smooth transition from the equilibrium radius to
the inflated radius, we let the donor inflate exponentially with
time, i.e. 
\begin{equation}
R_d \ = R_{d, {\rm inf}} \ = \ \left[ f_{\rm c} + ( 1 - f_{\rm c})e^{-t/10} \right] R_{d, {\rm eq}}, \label{EQ11}
\end{equation}
where $t$ in units of Myr is the time since the donor started 
to lose mass.
We assume that MB becomes inefficient when the donor mass reduces to 
0.2 M$_\odot$. This mass limit is significantly smaller than for single 
M dwarfs or M dwarfs in detached binaries ($0.3-0.35$ M$_\odot$) because 
the mass transfer above the gap drives the secondary out of thermal 
equilibrium \citep[see e.g.][for details]{Knigge_2006}.
At this point, the donor has time to 
relax to its thermal equilibrium radius. This can be described using an exponential ansatz \citep[see Appendix of][]{King_1995}.  
We use the following prescription \citep[see also][]{Davis_2008}:
\begin{equation}
R_d \ = \ \left[ 1 - ( 1 - f_{\rm max})e^{-t/10} \right] R_{d, {\rm inf}}, \label{EQ12}
\end{equation}
where $t$ (in units of Myrs) 
is the time since the donor detached from its Roche 
lobe and $f_{\rm max}~=~R_d/R_{d, {\rm CV}}^{\rm below}$ is the maximum 
inflation factor
with respect to the CV donor star radius below the gap, and 
$R_{d, {\rm CV}}^{\rm below}~=~0.225(M_d/0.2)^{0.61}$  \citep{Knigge_2011_OK}.
After the system has evolved as a detached binary through the gap, at an 
orbital period of $\approx$2 \,hrs, mass transfer starts again. We then again 
inflate the donor but to a lesser extent by using 
$R_{d, {\rm CV}}^{\rm above}~=~R_{d, {\rm CV}}^{\rm below}$ in Eq. \ref{EQ10}.
The same procedure, i.e. inflate the donor up to a radius given by $R_{d, {\rm CV}}^{\rm below}$, 
is applied for CVs that are born in or below the gap ($M_d < 0.35$ M$_\odot$).  
Finally, for period-bouncers ($M_d < 0.069$ M$_\odot$), we simply adopt 
$R_d~=~0.118(M_d/0.069)^{0.3}$ \citep{Knigge_2011_OK}.


\subsection{Angular momentum loss}
\label{aml}
The original version of \bse contains a prescription for disrupted
MB and takes into account AML through 
GR and tides \citep[][Sections 2.2,  2.3, and 2.4, 
respectively]{Hurley_2002}. 
In our upgraded version all procedures related to
AML due to stellar winds and tides remain the same as in
the original version. 
However, in the new version we provide an alternative MB prescription 
and normalization factors for MB and GR 
particularly suitable for CV evolution.  
We also add the possibility for additional AML that is 
generated by mass transfer, 
usually called consequential angular momentum loss 
(CAML, $\dot{J}_{\rm CAML}$)
\citep[e.g.][]{King_1995}.


\subsubsection{Consequential angular momentum loss}
\label{caml}

A potentially important ingredient for describing 
the evolution of close compact binary stars, in particular 
for CVs, is CAML which can significantly enhance 
the systemic AML generated by MB and GR.
It is clear that CAML exists in CVs as mass and also 
angular momentum is lost during nova eruptions. 
Recently, \citet{Schreiber_2016} and \citet{Nelemans_2016} also
discussed the possibility of 
additional AML due to frictional drag forces following 
nova eruptions. 
Alternative mechanisms for CAML in CVs are circumbinary disks 
\citep[e.g.][]{Willems_2005} or hydromagnetic accretion disk winds 
\citep[e.g.][]{Cannizzo_1988}. In low-mass X-ray binaries drag forces similar 
to those suggested by \citet{Schreiber_2016} and \citet{Nelemans_2016} 
might operate following X-ray bursts 
\citep{Gonzales_2017} 
as observed in V404 Cygni \citep{Munoz_2016}.

We have included two formulations for CAML in the revised {\sc bse} code, namely
the classical CAML (cCAML) from \citet{King_1995} and the empirical 
formulation (eCAML) from \citet{Schreiber_2016}. In cCAML it is
assumed that the material leaving the system during a nova eruption 
carries the specific angular momentum of the WD, while eCAML is
purely empirical and assumes stronger CAML 
for low-mass WDs to explain the absence of CVs with low-mass WD
primaries \citep{Zorotovic_2011}. 
The corresponding formulae we incorporated in the \bse code are: 
\begin{equation}
\frac{\dot{J}_{\rm CAML}}{J} \ = \ \nu \frac{\dot{M}_d}{M_d}, \label{EQ3}
\end{equation}
where
\begin{equation}
    \nu \ = \
  \begin{dcases}
    \frac{M_{d}^2}{M_{a}(M_{a} + M_{d})},	& \text{classical \citep{King_1995}} \\ 
				  	    	& \\
    \frac{0.35}{M_{a}}, 	  		& \text{empirical \citep{Schreiber_2016}} 
  \end{dcases} \label{EQ4}
\end{equation}
Both CAML prescriptions are optional in the upgraded 
{\sc bse} code and can be turned on/off if desired. However, if the user 
of \bse aims at simulating CV populations, assuming no CAML will likely lead to 
unrealistic results as mass and AML
during nova eruptions will be overlooked. 


\subsubsection{Magnetic braking (MB) and gravitational radiation (GR)}
\label{mbgr}

First of all, it is important to note that, in general, 
the strength and form of AML owing to MB is highly 
uncertain and we currently rely entirely on empirical relations. 
Systemic AML in the original {\sc bse} code is modeled as described in
\citet[][Section 2.4, Eqs. 48 and 50]{Hurley_2002}. 
This relation for MB 
seems to underestimate $\dot{J}_{\rm MB}$ in CVs as the resulting accretion
rates are significantly smaller than those derived from 
observations \citep{palaetal_2017}. 
Therefore, in the upgraded code, the equation for 
AML owing to MB was replaced with a prescription 
more commonly used to simulate CV evolution: 
\begin{equation}
\dot{J}_{\rm MB} \ = \ -5.83 \times 10^{-16}  M \left( R \Omega_{\rm spin} \right)^3  \ \ {\rm M}_\odot \, {\rm R}_\odot^2 \, {\rm yr}^{-2}, \label{EQ5}
\end{equation}
where $M$ and $R$ are the star mass and radius, respectively, both
in solar units, and $\Omega_{\rm spin}$ is the star spin in units of year
\citep{Rappaport_1983}.
Eq. \ref{EQ5} provides AML rates in better agreement with standard 
models of CV evolution and mass transfer rates similar to those
derived from observations. 
In the upgraded version of \bse the equation for AML 
due to GR remains the same as in the original. 

To reach agreement with the observations of CVs, it might be 
required to adopt normalization factors for both AML
prescriptions which strongly depend on the assumed CAML. 
If the classical CAML (or none) is assumed, 
then the normalization factors for MB and GR are 0.66 and 2.47 
\citep{Knigge_2011_OK}, 
respectively, and if the empirical CAML is adopted, they are 0.43 and 1.67 
\citep{Zorotovic_2016}, 
respectively.
We emphasize that these normalization factors are needed in order
to reconcile observed properties of CVs, such as mass transfer
rates, as well as the location of the orbital period minimum.
Whenever using the upgraded version of 
\bse one should thus keep in mind that these are purely empirical 
factors and that the physics behind them is far from understood.


\subsection{Stability of mass transfer}
\label{instcrit}

As explained by \citet{Schreiber_2016}, the stability of mass transfer
can be translated into a critical mass-ratio $q_{\rm crit} = M_d/M_a$ above 
which mass transfer is unstable (usually, dynamically or thermally). These
authors also showed that $q_{\rm crit}$ strongly depends on the adopted
form of CAML at the onset of mass transfer \citep[see also][]{King_1995}.

In the original {\sc bse} code, the critical mass ratio for mass transfer
on a dynamical time-scale is fixed to $q_{\rm crit}^{\rm dyn} = 0.695$ for
low-mass MS 
donors ($M \lesssim 0.8$ M$_\odot$), and no limit is applied 
for more massive MS stars (although $q_{\rm crit}^{\rm dyn} = 3$ was 
subsequently added to the code after publication). A more accurate way of determining 
$q_{\rm crit}^{\rm dyn}$ in CVs is to compare 
the adiabatic mass-radius exponent $\zeta_{\rm ad}$
of the MS star and the mass-radius exponent of the MS star 
Roche lobe $\zeta_{\rm RL}$. 
Therefore, in the upgraded {\sc bse} code 
we replaced the old criterion with the 
following.
For low-mass MS stars (i.e. $M_d \leq 0.8$ M$_\odot$), if $\zeta_{\rm RL} > \zeta_{\rm ad}$
then the CV is dynamically unstable, and stable otherwise. For more massive MS
stars, $q_{\rm crit}^{\rm dyn} = 3.0$. 
Note that $\zeta_{\rm RL}$ is given by \citet{Schreiber_2016} as 

\begin{equation}
\zeta_{\rm RL} \ = \ \frac{2}{3} \left( \frac{\ln(1+q^{1/3}) - \frac{1}{2}\frac{q^{1/3}}{1+q^{1/3}}}{0.6q^{2/3}+\ln(1+q^{1/3})}  \right) + 2\nu + \frac{M_d}{M_a+M_d} - 2, \label{EQ6}
\end{equation}
where $\nu$ is given by Eq.\ref{EQ4}. Alternatively, if CAML is not adopted, 
then \citep{Schreiber_2016} 
\begin{equation}
\zeta_{\rm RL} \ = \ \frac{2}{3} \left( \frac{\ln(1+q^{1/3}) - \frac{1}{2}\frac{q^{1/3}}{1+q^{1/3}}}{0.6q^{2/3}+\ln(1+q^{1/3})}  \right) (1+q) + 2(q - 1). \label{EQ7}
\end{equation} 
Finally, $\zeta_{\rm ad}$ is given by
\begin{equation}
    \zeta_{\rm ad} \ = \
  \begin{dcases}
    -\frac{1}{3},	& \text{if $ \ \ \ M_d \leq 0.38412$ M$_\odot$,} \\ 
				  	    	& \\
          f(M_d),	& \text{if $ \ \ \ 0.38412 < M_d/$M$_\odot$ $\leq$ 0.8,} \\ 
  \end{dcases} \label{EQ8}
\end{equation}
\
where $f(M_d) = 0.782 - 7.464M_d + 13.925M_d^2 - 5.389M_d^3$ 
\citep{Hjellming_1989}.
Similarly, we define a critical mass ratio $q_{\rm crit}^{\rm th}$ 
above which  mass transfer occurs on the 
thermal time-scale \citep{Hjellming_1989,Politano_1996}:

\begin{strip}
\begin{equation}
q_{\rm crit}^{\rm th} = \left\{
\begin{array}{rcl}
                                                         10.0566 M_d^2 - 4.54427 M_d + 1.60779,& \mbox{if} &  M_d \leq 0.25, \\
                                - 11.6147 M_d^3 + 14.8259 M_d^2 - 5.38702 M_d + 1.69848,& \mbox{if} &  0.25 < M_d \leq 0.58, \\
                                  23.9492 M_d^3 - 50.3456 M_d^2 + 34.3213 M_d - 6.35018,& \mbox{if} &  0.58 < M_d \leq 0.82, \\
                                - 17.2016 M_d^3 + 51.4274 M_d^2 - 49.7704 M_d + 16.8633,& \mbox{if} &  0.82 < M_d \leq 1.17, \\
- 2.01893 M_d^5 - 10.5146 M_d^4 + 109.538 M_d^3 - 278.879 M_d^2 + 289.253 M_d - 106.499,& \mbox{if} &  1.17 < M_d \leq 1.7, \\
                                                 0.185428 M_d^2 -0.552334 M_d + 1.35125,& \mbox{if} &  1.7 < M_d \leq 2.0, \\
                                                -0.123087 M_d^2 + 0.63407 M_d +0.211319,& \mbox{if} &  2.0 < M_d \leq 2.35, \\
                               0.00982875 M_d^3 -0.105005 M_d^2 +0.378526 M_d +0.585724,& \mbox{if} &  M_d > 2.35. \\
\end{array}
\right. \label{EQ9}
\end{equation}
\end{strip}

\begin{figure*}
   \begin{center}
    \includegraphics[width=0.49\linewidth]{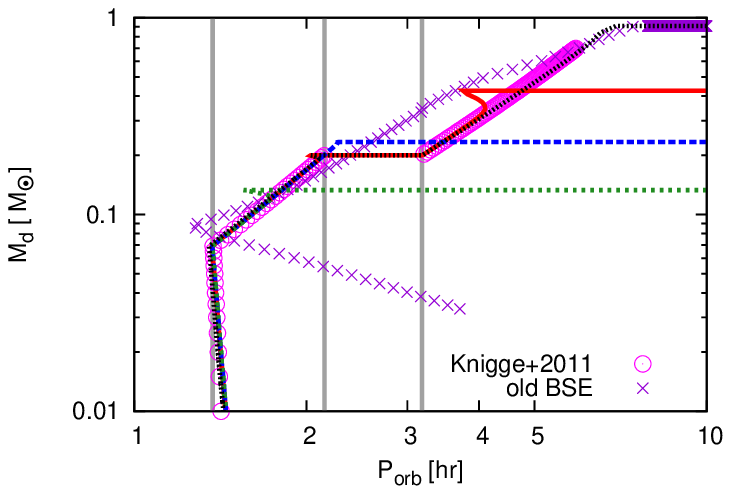} 
    \includegraphics[width=0.49\linewidth]{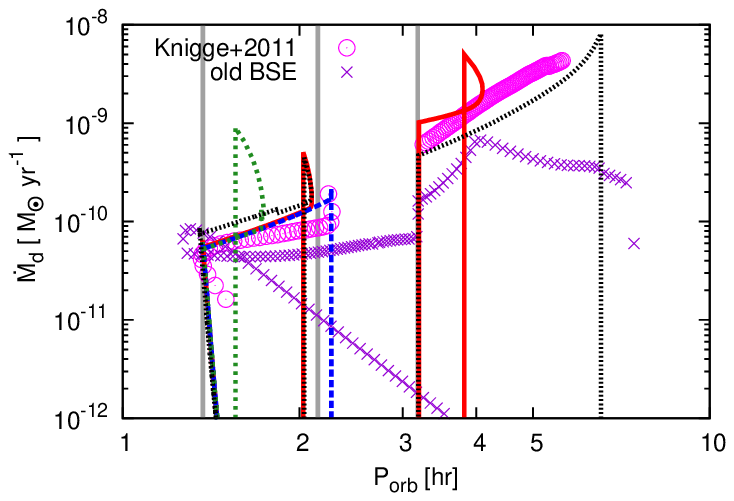} 
    \end{center}
  \caption{Evolution with orbital period $P_{\rm orb}$ of donor mass $M_d$
(left-hand panel), and mass transfer rate $\dot{M}_d$ (right-hand panel).
We show the evolution of four illustrative
CVs with initial donor and WD masses: 0.9 and 1.30 M$_\odot$ (dotted black line);
0.42 and 0.78 M$_\odot$ (solid red line); 0.25 and 0.73 M$_\odot$ (dashed blue line); and 0.13 
and 0.74 M$_\odot$(short dashed dark green line). 
We also include in the plots the tracks 
of the best-fit model by \citet{Knigge_2011_OK} (open pink circles) and the 
tracks of the original {\sc bse} code for a CV whose donor mass is 0.9 M$_\odot$ 
(dark violet crosses). For a better comparison with observational constraints,
we added to the plots the observational location of the period minimum \citep{Gansicke_2009} 
and gap edges \citep{Knigge_2006} (vertical gray lines). 
}
  \label{Fig01}
\end{figure*}

It is well known that PCEBs that start thermal time-scale mass 
transfer when the donor first fills its Roche-lobe 
may still become CVs with AML driven mass transfer as soon as the 
mass ratio becomes $\lappr\,1$ \citep{schenkeretal02-1}
and if they do not become dynamically unstable first. 
Such CVs and their progenitors have indeed been observed 
\citep{gaensickeetal03-1,parsonsetal15-1} but most likely do not make up 
more than a few per cent of the entire CV population. 
The original prescription from 
\citet{Hurley_2002} is known to underestimate the mass transfer rates for
thermal time-scale mass transfer.  
This problem can be fixed to some degree by incorporating 
an additional factor in the mass transfer equation 
but the obtained mass transfer can still be off by more than a factor of 
three \citep{Claeys_2014,Wijnen_2015}. 
For simplicity and because only few CVs are assumed to have evolved through
thermal time-scale mass transfer, we kept the original 
thermal time-scale mass transfer equation in the upgraded version 
of \bse. The only difference is that
the procedure described in \citet[][sections 2.6.3 and 2.6.6]{Hurley_2002} is 
applied only if the CV is going through thermally unstable mass transfer. 
Users of the upgraded version of \bse should take into account 
the potential shortcomings of the treatment of thermal time-scale mass transfer.


\subsection{Comparison with old prescriptions}
\label{compoldnew}

In order to ilustrate that our modifications lead to CV evolution that
is in much better agreement with observations than the old prescriptions,
we discuss the evolution of four individual binaries with 
MS donor masses of 0.13, 0.25, 0.42 and 0.9 M$_\odot$
and WD masses of 0.74, 0.73, 0.78 and 1.30 M$_\odot$, respectively. 
According to the masses of the donor stars, two of them are born above the gap
(i.e. long-period CVs)
and one each below (i.e. short-period) and in the gap. Figure\,\ref{Fig01} shows the
evolution of these systems in the 
donor mass vs. period (left-hand panel) plane and mass 
transfer rate vs. period (right-hand panel) plane. 
In addition to the predictions of the upgraded and the original 
{\sc bse} code we also show the best-fit model 
of \citet{Knigge_2011_OK}.

Figure \ref{Fig01} clearly shows that the period gap cannot be reproduced by 
the original {\sc bse} code (dark violet crosses in Fig. \ref{Fig01}).
While the mass transfer rate drops when the donor becomes fully convective
because AML from MB vanishes (at an orbital period of $\approx3\,$hrs), 
the system born above the gap never detaches because the radius of 
the donor is assumed not to be affected by mass transfer. The second obvious flaw of the original \bse code 
are the extremely long periods predicted for period-bouncers
which are caused by assuming a non-realistic donor mass-radius 
relation when the donor becomes degenerate (Section \ref{donor}). 
In addition, the mass transfer
rate predicted by the original code is not in agreement 
with the model derived from observations (pink circles in Fig. \ref{Fig01}), 
especially above the gap. This is caused by underestimating the
strength of MB, as discussed in Section \ref{mbgr}. 

In contrast, the predictions for CV evolution 
obtained with the upgraded code (lines in Fig. \ref{Fig01}),
are in agreement with the observations. 
The orbital period gap position and width, as well as the location of the 
period minimum are reproduced correctly. 
In addition, period-bouncers stay close to the period minimum with small mass
transfer rates and the mass transfer rates predicted for CVs above the
gap are more realistic. 

If fact, let us take as an example the system with WD and 
donor masses 0.42 and 0.78 M$_\odot$, 
respectively, which is indicated by the solid red line. 
We first note that 
this system's orbital period decreases due to MB as long as it remains
detached (horizontal part of the evolutionary track) 
and starts 
its life as a long-period CV at $\approx$ 3.5 hrs, when mass transfer starts. At this
point, the donor is driven out of thermal equilibrium 
and becomes significantly bloated which causes 
an increase in the orbital period. When reaching the
CV donor mass-radius relation, the orbital period starts to decrease until MB braking
becomes inefficient, i.e. when the donor becomes fully convective ($M_d \approx 0.2$ M$_\odot$)
at a period of $\approx$ 3 hrs. At this period, the donor has time to relax to
its thermal equilibrium radius and the systems becomes detached. As GR is still
acting to remove angular momentum, the orbital period 
keeps decreasing. This phase of detached 
evolution corresponds
to the orbital period gap (vertical line between 2 
and 3 hrs).
When the orbital period is $\approx$ 2 hrs,
the Roche lobe has shrunk enough to restart mass 
transfer and the
system becomes a CV again 
(now as a short-period system). The orbital period 
is further decreasing until the 
moment the 
the thermal time-scale exceeds
the mass loss time-scale and the donor starts 
expanding in response to the mass loss. This
happens when $M_d \approx 0.07$ M$_\odot$ and $P_{\rm orb} \approx 1.4$ hrs.
From this point on, the CV is a period-bouncer 
and the donor is degenerate,
which causes the period to slightly increase 
in response to further mass loss and AML.


\subsection{Final Considerations}
\label{finalc}

We have presented an upgrade of the \bse code that is suitable for performing 
binary population synthesis of CVs. 
This upgraded version contains new AML 
and mass transfer prescriptions.
The results we obtain using the upgraded code are in agreement with the
standard model of CV evolution and the observations. Some of 
the prescriptions we incorporated in the new version of \bse 
are empirical rather than being theoretically well understood. Examples are the 
strength of magnetic
braking which is normalized to reproduce the mass transfer rates of CVs above
the period gap, the empirical CAML prescription from
\citet{Schreiber_2016} which has been invented to explain the absence of
low mass WDs among CV primary stars, the mass radius relation for CV donors 
above the period gap which has been derived from observations \citep{Knigge_2011_OK}, 
and 
the mass of the donor star at which MB becomes inefficient 
which has been adjusted to reproduce the width of the orbital period gap 
\citep{Knigge_2011_OK}. Users should be aware of the empirical nature of these 
prescriptions when using the 
upgraded \bse code. 

Another important fact to be considered is the increased runtime of the
upgraded code.  
The new mass transfer prescription requires smaller timesteps 
($\lappr 10^{3-4}$ yrs)
which makes the code significantly slower. 
The original {\sc bse} code evolves 1 million binaries 
in $\sim$ 3 hrs while
the upgraded code needs $\sim$ 1 day to evolve the same amount of 
binaries\footnote{The 
simulations presented in this paper 
were performed on a PSK cluster at the Nicolaus Copernicus 
Astronomical Centre in Poland.}. 
However, the upgraded code is still the fastest CV evolution code 
available and is therefore perfectly suitable for population synthesis 
of CVs.
Indeed, we are aware of only one code publicly   
available 
that is able to correctly reproduce CV
evolution, i.e. the {\sc mesa} code 
\citep[e.g.][]{Paxton_2011,Paxton_2015,Kalomeni_2016}. 
This stellar evolution code is more 
accurate than \bse as it does not rely
on analytic approximations, but it is also orders 
of magnitude slower. 
Other fast publicly 
available population synthesis codes are {\sc SeBa} 
\citep[e.g.][]{Toonen_2012} and {\sc binary\_c} \citep[e.g.][]{Izzard_2006}. 
These codes are, however, unable to reproduce 
CV evolution 
as described here
and give similar results to the old \bse code, i.e. 
they are similarly not suitable for performing detailed CV population 
studies (although the upgrades described here can be incorporated into those codes).

In addition to being somewhat slower, 
CV evolution according to 
the upgraded version of \bse is limited by assuming a 
priori solar metallicities. 
It is still possible to use different metallicities 
in the input file of the
upgraded code (in the same way as in the original 
code), 
but the metallicity parameter only affects pre-CV evolution such as the 
lifetime of the WD progenitor on the MS.  
As soon as a WD+MS binary starts stable AML driven mass transfer 
metallicity effects are ignored as we simply use the mass radius relation of
\citet{Knigge_2011_OK} for the donor star. 
Fortunately, for population models of CVs metallicity effects are expected to be very small. 
For example, \citet{Stehle_1997} compared CV evolution for solar and 
sub-solar metallicity and find that low-metallicity CVs 
produce a slightly smaller period gap, a somewhat shorter
minimum period, and in general slightly higher mass transfer
rates. This leads to shorter evolutionary time-scales compared to
CVs with donors having a solar chemical composition but otherwise the overall
evolution is identical. Thus, even though such small differences exist, 
they should not drastically affect the results of population synthesis, since
the objective in such investigations is not to model particular CVs, 
but rather to establish an overall (statistical) picture for CV evolution.

We would like to draw the readers attention to one important 
but often overlooked or misinterpreted improvement 
of the {\sc bse} code that was incorporated after the publication of the
original {\sc bse} paper \citep{Hurley_2002}. CEP can 
now be calculated taking into account recombination 
energy as described in \citet[][see their Appendix A]{Claeys_2014}. 
The efficiency of recombination energy in contributing to expelling the
envelope can be controlled with the parameter $\lambda$. 
A positive value of $\lambda$ represents the fraction of the 
recombination energy 
included in the calculation of the binding energy parameter, 
$\lambda = 0$ implies that the binding energy parameter is computed assuming
recombination energy does not
contribute, while for $\lambda < 0$, the 
binding energy parameter is fixed and set equal to -$\lambda$ 
\citep[see also][]{zorotovicetal14-2}. 

Finally, all parameters in the code are listed, explained 
and defined in a file called \texttt{parameters.h}. 
The source code contains an 
example for population synthesis of CVs and can be downloaded 
from \href{http://www.ifa.uv.cl/BSE}{\texttt{http://www.ifa.uv.cl/BSE}} or 
\href{http://astronomy.swin.edu.au/~jhurley}{\texttt{http://astronomy.swin.edu.au/\~{}jhurley}}.

\bsp	
\label{lastpage}
\end{document}